\documentstyle[12pt,epsf]{article} 

\renewcommand{\thefootnote}{\fnsymbol{footnote}}

\begin{document}


\begin{titlepage}
\begin{flushright}
\begin{tabular}{l}
NORDITA--97--59 P\\
UAB--FT--425\\
hep-ph/9709243
\end{tabular}
\end{flushright}
\vskip0.5cm
\begin{center}
  {\Large \bf Radiative Corrections to the Decay $B\to \pi e\nu$ and
   the Heavy Quark Limit\\}
  \vskip1cm {\large E. Bagan}
  \vskip0.2cm
  Grup de F\'\i\/sica Te\`orica, Dept.\ de F\'\i\/sica and Institut de 
  F\'\i\/sica d'Altes Energies, IFAE, Universitat Aut\`onoma de Barcelona,
  E--08193 Bellaterra, Spain\\ 
  \vskip1cm {\large Patricia Ball}
  \vskip0.2cm
  Fermi National Accelerator Laboratory,
  P.O.\ Box 500, Batavia, IL 60510, USA\\
  \vskip1cm {\large V.M.\ Braun}\footnote{On leave of absence from
    St.\ Petersburg Nuclear Physics Institute, 188350 Gatchina, Russia.}\\
  \vskip0.2cm
  NORDITA, Blegdamsvej 17, DK--2100 Copenhagen, Denmark\\
  \vskip3cm
{\large\bf Abstract:\\[10pt]} \parbox[t]{\textwidth}{ 
  We calculate radiative corrections to the light-cone sum rule
  for the semileptonic form factor in $B\to\pi e\nu$ decays and thus
  remove the major uncertainty in determining the CKM mixing angle
  $|V_{ub}|$ by this method.
We discuss  the remaining uncertainties and perspectives for
  further studies. 
  The structure of the radiative corrections
   suggests factorization of soft (end-point)  and 
   hard rescattering contributions in
   heavy-to-light decays in the heavy quark limit.} 
  \vskip1cm 
{\em Submitted to Physics Letters B}
\end{center}
\end{titlepage}

\renewcommand{\thefootnote}{\arabic{footnote}}
\setcounter{footnote}{0}



\noindent {\large\bf 1.} 
Precise determination of the CKM mixing angle $|V_{ub}|$ will most likely 
come from semileptonic $B$ decays. Two competing strategies exist, the
study of the end-point region in inclusive decays and the analysis
of various
exclusive  decay channels, most notably $B\to\pi e\nu$ which is the easiest
one to access 
experimentally. The theoretical challenge in the latter case is to calculate
the $B\to\pi$ transition form factor $f_+(q^2)$ induced by the weak vector
current:
\begin{equation}
  \langle \pi(p_\pi)| \bar u\gamma_\mu b |B(p_B)\rangle =
   (p_B+p_\pi)_\mu\, f_+(q^2) + (p_B-p_\pi)_\mu f_-(q^2)\,.
\label{def:f+}
\end{equation} 
Here $q^2 = (p_B-p_\pi)^2$ is the square of the momentum 
 transferred to the lepton
pair; the second form factor $f_-$ does not contribute to the decay rate 
for zero mass leptons and will not be considered in this letter.
 The decay rate is dominated by contributions from small
values of $q^2$, for which the $u$ quark produced in the weak decay
has large energy of order 
$m_b/2$ in the $B$ meson rest frame ($m_b$ is the $b$ quark pole mass). A consistent QCD
description of such  processes  exists up to now only in the Sudakov limit \cite{ASY94}, in which
case contributions from large transverse quark-antiquark separations 
are suppressed and the form factor is dominated by hard gluon exchange 
\cite{hard}. This limit is theoretically interesting, but not relevant for 
realistic $b$ quark masses. In fact, we will argue that ``soft''
contributions related to large transverse distances exceed the ``hard''
contributions by an order of magnitude (and have opposite sign).
Any theoretical approach to heavy-to-light decay form factors aiming at
quantitative predictions has to deal with these ``soft'' terms explicitly,
as do e.g.\ light-cone sum rules and various models~\cite{EPStalk}.

The consistent separation of ``hard'' and ``soft'' contributions in 
heavy-to-light
decays presents an unsolved problem which, to our knowledge, has never 
been tackled in a systematic way\footnote{One reason being that in 
the classical application, the pion form factor, the ``soft'' terms are 
suppressed by one power of the momentum transfer.}. In this letter
we examine a possibility
for such a separation, based on our calculation of the 
radiative corrections to the corresponding light-cone sum rules. 

The basic idea of the light-cone sum rule approach is to consider a two-point 
correlation function replacing the $B$ meson state by a suitable interpolating
current:
\begin{eqnarray}
 \Pi_\mu(p_B^2,q^2) &=& i\int d^4x\, {\rm e}^{-ip_Bx}
\langle \pi(p_\pi)|
T\{\bar u(0)\gamma_\mu b(0)\bar b(x)i\gamma_5 d(x)\}|0\rangle
\nonumber\\
&=& (p_B+p_\pi)_\mu\, \Pi_+(p_B^2,q^2) + (p_B-p_\pi)_\mu\, \Pi_-(p_B^2,q^2). 
\label{def:Pi_mu}
\end{eqnarray}
The Lorentz-invariant function $\Pi_+$ has a pole at $p_B^2=m_B^2$ corresponding
to the contribution of the $B$ meson:
\begin{equation}
 \Pi_+(p_B^2,q^2) = 
\frac{f_B m_B^2}{m_b} \frac{f_+(q^2)}{m_B^2-p_B^2}+\ldots,   
\label{lhs}
\end{equation}
where $m_B$ is the $B$ meson (pole) mass and the dots stand for contributions from higher-mass
resonances and the  continuum. The $B$ meson decay constant $f_B$ is usually defined as
\begin{equation}
m_b \langle 0|\bar d\,i\gamma_5 b |B\rangle = m_B^2 f_B.
\label{def:fB}
\end{equation}

On the other hand, the correlation functions in (\ref{def:Pi_mu}) can 
be calculated in the Euclidean region, $p^2_B-m_B^2$ large and negative,
using the 
light-cone expansion. Up to higher twist corrections 
the product of $b$ quark fields can be substituted by a perturbative 
propagator and the tree-level result is 
\begin{equation}
  \Pi_+(p_B^2,q^2) = 
 \frac{1}{2}
  f_\pi m_b \int_0^1 du \frac{\phi_\pi(u,\mu)}{m^2_b-u p_B^2 -\bar u q^2}.
\label{rhs}
\end{equation} 
Here and below $\bar u = 1-u$, $\phi_\pi(u,\mu)$ is
the leading twist pion
distribution amplitude defined by
\begin{equation}
\langle \pi(p_\pi)|\bar u(0)\gamma_\nu\gamma_5 d(x)|0\rangle
\stackrel{x^2=0}{=} -i f_\pi p_{\pi\nu} 
\int_0^1 du\, {\rm e}^{i\bar up_\pi x}\phi_\pi(u,\mu)
\label{def:phi_pi}
\end{equation}
in Fock-Schwinger gauge and
$\mu$ is the factorization (renormalization) scale. The variable $u$ has the 
physical meaning of the momentum fraction carried by the $u$ quark in
the infinite momentum frame.

Next, we use the concept of duality, assuming that the contribution of
the $B$ meson 
to (\ref{lhs}) corresponds to an integral over the spectral 
density calculated within the light-cone expansion, Eq.~(\ref{rhs}),
in a certain duality interval:
\begin{equation}
\Pi_+^{\rm B\ meson}(p_B^2,q^2) = 
\int_{m_b^2}^{s_0}\frac{ds}{s-p_B^2}\,\rho(s,q^2).
\label{duality}
\end{equation}
The width of the duality interval is characterized by the parameter 
$s_0$ (continuum threshold) and in general is of order $s_0-m_b^2\sim O(m_b)$.
The spectral density $\rho(s,q^2)$ immediately 
follows from Eq.~(\ref{rhs}) after a simple change of variables 
$s\to (m^2_b-q^2)/u+q^2$. Thus, in this particular case
the duality restriction on the maximum invariant
energy $s_0$ translates into a restriction on the {\em minimum} momentum
fraction $u_0$ carried by the $u$ quark:
\begin{equation}
  u_0 = \frac{m_b^2-q^2}{s_0-q^2}.
\label{def:u0}
\end{equation}
 Equating the representations (\ref{lhs}) and (\ref{rhs}) and
 isolating the contribution of the $B$ meson according to
 (\ref{duality}), we obtain the simplest version of the
light-cone sum rule for $f_+(q^2)$, which neglects higher twist and
 radiative corrections. In order to suppress the contributions of
 higher order states, it is written in a Borel
transformed form, which amounts to replacing the factor $1/(s-p_B^2)$
 by $\exp(-s/M^2)$ and
$1/(m_B^2-p_B^2)$ by $\exp(-m_B^2/M^2)$:
\begin{equation}
  \frac{f_B m^2_B}{f_\pi m^2_b}\,{\rm e}^{(m_b^2-m_B^2)/M^2} f_+(q^2) =
   \frac{1}{2}
   \int_{u_0}^1 du\, \frac{\phi_\pi(u,\mu)}{u}\,{\rm e}^{-\bar u 
(m^2_b-q^2)/(uM^2)}.
\label{SR:tree}
\end{equation}
Here $M^2$ is the Borel parameter. 


\bigskip

\noindent {\large\bf 2.} 
The accuracy of the sum rule (\ref{SR:tree}) can be improved by including
higher twist and radiative
corrections. The former ones have been calculated earlier 
\cite{BKR93,Bel95}; Eq.~(79) in \cite{Bel95} gives the complete result 
to twist 4 accuracy, which we will use in the numerical analysis.
We have calculated first order radiative corrections to the leading
twist term (\ref{SR:tree})
 The calculation is straightforward and similar to 
earlier calculations 
of the radiative correction to the
$\gamma^*\to\pi\gamma$ form factor \cite{Braaten}, but more cumbersome 
because of the nonzero $b$ quark mass. In the $\overline{\rm
  MS}$-scheme, the result reads:
\begin{eqnarray}
\lefteqn{
\frac{f_B m^2_B}{f_\pi m^2_b}\,{\rm e}^{(m_b^2-m_B^2)/M^2} f_+(q^2)\,\, =\,\,
\frac{1}{2}   
\int_{u_0}^1 du \frac{\phi_\pi(u,\mu)}{u}\,{\rm e}^{-\bar u (m^2_b-q^2)/(uM^2)}}
\nonumber\\
 & - & C_F \frac{\alpha_s}{4\pi}\, {\rm e}^{m_b^2/M^2}
\left\{
\int_0^1 {\rm d}u\, {\phi_\pi(u,\mu)\over\bar u}\int_{m_b^2}^{s_0}
{\rm d}t\,{m_b^2-t\over t-q^2}\left({2\over t}+L_1\right) {\rm e}^{-t/M^2}
\phantom{\left[{{\rm e}^{-{ut+\bar u q^2-m_b^2\over uM^2}}
   -1\over m_b^2-ut-\bar u q^2}\right]}
 \right. \nonumber \\
&+&
\int_0^{u_0} {\rm d}u\, 
 {\phi_\pi(u,\mu)\over \bar u}{\rm e}^{-{m_b^2-\bar u q^2\over uM^2}}
\int_{m_b^2}^{s_0}
{\rm d}t\,(m_b^2-t) {{\rm e}^{-{ut+\bar u q^2-m_b^2\over uM^2}}
   \over m_b^2-ut-\bar u q^2}\;L_2\;
\nonumber\\
&+&\int_{u_0}^1 {\rm d}u\, 
{\phi_\pi(u,\mu)\over\bar u}{\rm e}^{-{m_b^2-\bar u q^2\over uM^2}}
\left[
\int_{m_b^2}^{s_0}
{\rm d}t\,(m_b^2-t){{\rm e}^{-{ut+\bar u q^2-m_b^2\over uM^2}}-
 1\over m_b^2-ut-\bar u q^2}\;L_2
+\int_{m_b^2}^{s_0}
{\rm d}t\,L_2\right.\nonumber\\
&&
\phantom{\left.\int_{u_0}^1 {\rm d}u\, 
{\phi_\pi(u,\mu)\over\bar u}{\rm e}^{-{m_b^2-\bar u q^2\over uM^2}}\right]}
\kern-1.0cm
\left.
-\int_{m_b^2}^{t_0}
{\rm d}t\,{m_b^2-q^2\over t-q^2}\left[
{m_b^2-t\over t(m_b^2-q^2)}+L_1
\right] {\rm e}^{-{t-m_b^2\over uM^2}}
\phantom{{{\rm e}^{-{ut+\bar u q^2-m_b^2\over uM^2}}-1
\over m_b^2-ut-\bar u q^2}}\kern-2.3cm
\right]\nonumber\\
&+&
\int_{u_0}^1 {\rm d}u\, 
{\phi_\pi(u,\mu)\over u}{\rm e}^{-{m_b^2-\bar u q^2\over uM^2}}
\left[{5\over2}
-{\gamma_E\over2}+2\ln{u M^2\over m^2_b}-{3\over2}
\ln{u M^2\over \mu^2}+{1\over2}{\rm Ei}\left({m_b^2-t_0\over u M^2}\right)
\phantom{{{\rm e}^{-{ut+\bar u q^2-m_b^2\over uM^2}}-1
\over m_b^2-ut-\bar u q^2}}\kern-2.3cm
\right.\nonumber\\
&&
\phantom{\left.\int_{u_0}^1 {\rm d}u\, 
{\phi_\pi(u,\mu)\over u}{\rm e}^{-{m_b^2-\bar u q^2\over uM^2}}\right]}
\kern-1.0cm
+\int_{m_b^2}^{s_0}
{\rm d}t\,L_2
+\int_{s_0}^{\infty}
{\rm d}t\,{(m_b^2-q^2)\;L_2\over m_b^2-ut-\bar u q^2}
+
\int_{m_b^2}^{t_0}
{\rm d}t\,\left({1\over t}-L_2\right){\rm e}^{-{t-m_b^2\over uM^2}}
\nonumber\\
&&
\phantom{\left.\int_{u_0}^1 {\rm d}u\, 
{\phi_\pi(u,\mu)\over u}{\rm e}^{-{m_b^2-\bar u q^2\over uM^2}}\right]}
\kern-1.0cm
+\int_{m_b^2}^{t_0}
{\rm d}t\,\left[
{1\over2}-{(t-m_b^2)^2\over2t^2}+(m_b^2-q^2)\left(L_1+L_2\right)
\right]{{\rm e}^{-{t-m_b^2\over uM^2}}-1\over m_b^2-t}\nonumber\\
&&
\phantom{\left.\int_{u_0}^1 {\rm d}u\, 
{\phi_\pi(u,\mu)\over u}{\rm e}^{-{m_b^2-\bar u q^2\over uM^2}}\right]}
\kern-1.0cm
\left.\left.
-\int_{t_0}^{\infty}
{{\rm d}t\over m_b^2-t}\,\left[
{1\over2}-{(t-m_b^2)^2\over2t^2}+(m_b^2-q^2)\left(L_1+L_2\right)
\right]
\phantom{{{\rm e}^{-{ut+\bar u q^2-m_b^2\over uM^2}}
-1\over m_b^2-ut-\bar u q^2}}\kern-2.3cm
\right]
\right\}\,,
\label{SR:rad}
\end{eqnarray}  
where $m_b$ is the one-loop pole mass, 
${\rm Ei}(x)$ the exponential-integral function, defined as 
${\rm Ei}(x)=-\int_{-x}^\infty {\rm d} y\; {\rm e}^{-y}/y$, 
$t_0 \equiv u s_0 + \bar u q^2$ and 
\begin{eqnarray}
&\displaystyle
L_1={1\over t-q^2}\left[-1+\ln{(t-m_b^2)^2\over t\mu^2}\right];\qquad
L_2={1\over t-q^2}\left[-{m_b^2\over t}+\ln{(t-m_b^2)^2\over t\mu^2}\right].&
\end{eqnarray}
Details of the calculation will be published elsewhere. We have
checked that the $\mu$ dependent terms 
cancel the $\mu$ dependence of the distribution amplitude
$\phi_\pi(u,\mu)$ to leading logarithmic accuracy \cite{exclusive}.
We have also 
checked that our results agree with the recent results of Ref.~\cite{KRWY},
where a similar calculation is reported, both   
analytically for the amplitude in momentum space and 
numerically for the final sum rule after Borel
transformation and continuum subtraction. Our representation for the 
final answer is simpler.

\bigskip 

\noindent {\large\bf 3.}
In order to gain a better understanding of the structure of
radiative corrections, it
is convenient to go to the heavy quark limit $m_b\to\infty$.
In this limit the dimensionful parameters
$f_B$, $s_0$ and $M^2$ have to be rescaled as
\begin{eqnarray}
f_B = 1/\sqrt{m_b}\,f^{\rm stat}(\mu=m_b), &\qquad& m_B-m_b = \bar\Lambda,
\nonumber\\
s_0 = m_b^2 + 2 m_b \omega_0, &\qquad&
M^2 = 2 m_b \tau,
\end{eqnarray} 
where $\omega_0$ and $\tau$ are the nonrelativistic continuum threshold
and the Borel parameter, respectively. The heavy quark limit of the form factor
depends crucially on the value of $q^2$. As we are  only interested in small values
$q^2\ll m_b^2$, we will set $q^2=0$ in this discussion for simplicity.

In this limit the momentum fraction cut-off (\ref{def:u0}) becomes 
$u_0 = 1- 2\omega_0/m_b$, so that the integration region in (\ref{SR:tree})
shrinks to a narrow interval near the end-point, corresponding to all the
pion momentum being carried by the $u$ quark. Since close to the end-point 
$\phi_\pi(u,\mu)\sim(1-u)$, we obtain
\begin{equation}
  \frac{f^{\rm stat}(\mu=m_b)}{\sqrt{m_b}f_\pi} {\rm e}^{-\bar\Lambda/\tau}f_+(0)
 = -\frac{2}{m_b^2}\phi_\pi'(1)\int_0^{\omega_0} d\omega\, 
\omega {\rm e}^{-\omega/\tau}
\label{SR:hqltree}
\end{equation}
where $\phi_\pi'(u) = (d/du)\phi_\pi(u)$. Note that
$\phi_\pi'(1)<0$. One thus finds
$f_+(0)\sim m_b^{-3/2}$, at least at tree level \cite{CZ90,ABS}.

The full expression for the radiative corrections 
looks rather complicated; to make its structure more transparent
we take the limit $\tau\to\infty$, corresponding to the so-called local duality 
approximation:
\begin{eqnarray}
\lefteqn{
    \frac{f^{\rm stat}(m_b)}{f_\pi}[m_b^{3/2}f_+(0)]=}
\nonumber\\
    &=&-\omega_0^2\phi_\pi'(1)\Bigg[ 1+ \frac{\alpha_s}{\pi}C_f
    \Bigg(\frac{1+\pi^2}{4} +\ln\frac{m_b}{2\omega_0}  
   -\frac{1}{2}\ln^2\frac{m_b}{2\omega_0} 
    +\frac{1}{2}\ln \frac{2\omega_0}{\mu}
    \Bigg)\Bigg]  
\nonumber\\
&&{} -\omega_0^2 \frac{\alpha_s}{\pi}C_f
\left[ 
     \left(1-\ln\frac{2\omega_0}{\mu}\right)
   \int_0^1 du \left(\frac{\phi_\pi(u)}{\bar u^2}+
\frac{\phi_\pi'(1)}{\bar u}\right)
     - \ln\frac{2\omega_0}{\mu}\int_0^1 du \,\frac{\phi_\pi(u)}{\bar u}
\right].
\label{SR:hqlrad}
\end{eqnarray}  
This expression deserves to be studied in some detail.
Let us interpret the two pieces:
the first term on the right-hand side must be identified with the 
soft (end-point) contribution including the Born-term and its
radiative correction, while the second term corresponds to 
the usual mechanism of hard gluon exchange.

The dependence on the collinear factorization scale $\mu$
must cancel the scale dependence of the pion distribution amplitude.
This implies that the structure of terms in $\ln \mu$ 
in the hard contribution is fixed by the structure of the leading order
soft term which is proportional to $\phi_\pi'(1,\mu)$.
Indeed, we find
\begin{eqnarray}
   \frac{d}{d \ln \mu} \phi_\pi'(1,\mu) &=& \frac{\alpha_s}{\pi}C_f
   \frac{d}{dx}\left[\int_0^1 dy\, V_0(x,y)\,\phi_\pi(y,\mu)\right]_{x\to 1} 
\nonumber\\
&=&{}-\frac{\alpha_s}{\pi}C_f\left\{
      \int_0^1 du\,\left[\frac{\phi_\pi(u)+\bar u \phi_\pi'(1)}{\bar u^2} +
      \frac{\phi_\pi(u)}{\bar u}\right]-\frac{1}{2}\phi_\pi'(1)
                             \right\},
\label{scaledepend}
\end{eqnarray}
where $V_0(x,y)$ is the usual Brodsky-Lepage kernel, so that
the structure of $\ln \mu$ terms in (\ref{SR:hqlrad}) is reproduced. 
Note the subtraction term accompanying the naively divergent expression
$\int du\, \phi_\pi(u,\mu)/\bar u^2$ \cite{hard}, which is similar to the
usual  ``plus'' prescription in the evolution kernel.

Local duality means that we identify the $B$ meson with
a $b$ quark accompanied by an arbitrary number of light quarks and gluons 
with total energy less than $\omega_0$ (in the $b$ quark rest frame).
Consider the ``deep inelastic'' cross section of neutrino scattering
off a pion, in which one selects the contribution of the charged weak current
with a $b$ quark in the final state: 
$d\sigma /d M^2_X(\nu_e+\pi \to X_b+ e)$,
where $M_X^2$ is the invariant mass of the hadronic final state. 
In the approximation adopted in this letter we identify the integral 
of this cross section over the region of small invariant masses 
$M^2_X < m_b^2+2 m_b \omega_0$ with the square of the form factor for
the inverse process $F(\nu\pi\to e B)$ (up to kinematical factors). 
This interpretation is useful in several aspects.
It is easy to check that the ``soft'' contribution in (\ref{SR:hqlrad})
corresponds to the would-be leading twist contribution to the 
deep inelastic cross section, while the ``hard'' contribution 
(involving interaction with the quark-spectator) is of higher twist.
These two terms are of the same order (in the $b$ quark mass) since 
the leading twist contribution is additionally suppressed by a factor 
$1/m_b^2$ for small values of the invariant mass of the 
hadronic system, because the corresponding parton
distribution (in the pion, for the case at hand) vanishes 
at the end-point $x\to1$. 
With a low collinear factorization scale 
$\mu\sim \omega_0$ all quark mass dependence in (\ref{SR:hqlrad}) is due to
the ``soft'' contribution and thus to the leading twist part of the 
cross section. This suggests that resummation of heavy quark mass 
logarithms can be done using the same techniques as for the end-point 
spectrum in inclusive $b\to u$ decays \cite{KS94}. A detailed discussion
goes beyond the scope of this letter. 

A final remark concerns the size of the radiative correction. 
With the natural factorization scale $\mu = 2 \omega_0\simeq 2$~GeV
and with $m_b\simeq 5$~GeV the quark mass logarithms are 
of order unity and the large constant term dominates. It has to be compared,
however, to the large radiative correction which was found in QCD sum rules
for the decay constant $f^{\rm stat}$. In the same (local duality)
approximation and neglecting contributions of condensates,
one finds \cite{BBBD}
\begin{equation}
   f^{\rm stat}(m_b) = \frac{\omega_0^{3/2}}{\pi}
   \left[ 1 + \frac{\alpha_s}{\pi}C_f
   \left( \frac{15}{8}+\frac{1}{6} \pi^2 +\frac{3}{4}
   \ln\frac{m_b}{2\omega_0}\right)
   \right].
\end{equation}
We see that the large radiative corrections almost cancel each other 
between $f^{\rm stat}$ and the right-hand side of (\ref{SR:hqlrad}). 
This shows
that the form factor itself is free from large radiative corrections
in the light-cone sum rule approach.  
The same cancellation takes place
in the complete expressions with finite $b$ quark mass.

\bigskip

\noindent {\large\bf 4.}
We have carried out a detailed numerical analysis of the complete 
sum rules with finite $b$ quark mass and including radiative and
higher twist corrections. To this end we substitute the value of $f_B$ in
(\ref{SR:rad}) by the corresponding sum rule \cite{AE} including
radiative corrections with the same 
value of the continuum threshold and the same Borel parameter. 
In particular, we use the range
$5\,{\rm GeV}^2<M^2<8\,{\rm GeV}^2$ and
$m_b=\{ 4.6, 4.7, 4.8 \}\,{\rm GeV}$
with the continuum thresholds
$s_0=\{ 35,34,33.5\}\,{\rm GeV}^2$, respectively. 
The resulting value for $f_B$ is
$(175\pm 25)\,{\rm MeV}$, which is in  agreement with the
current lattice average \cite{Flynn96}. 
We do not attempt a renormalization group improvement (resummation of 
$\ln m_b$) of the sum rules and set the factorization scale to 
$\mu = \sqrt{m_B^2-m_b^2}\simeq 2.4$~GeV. The scale dependence of our
results is in fact negligible.
A remark is in order about the actual
choice of $M^2$ in Eq.~(\ref{SR:rad}). 
The expansion turns out to be essentially in inverse powers of
$u M^2$ rather than $M^2$ itself. In order to avoid small values of this
``effective'' parameter $u M^2$ for large $q^2$ (where $u$ can become
small), we use the rescaled value $M^2=M^2_{2pt}/\langle u\rangle$ in
the Borelized expression for the correlation function $\Pi_+$ and
$M^2_{2pt}$ in the above mentioned range in the sum rule for
$f_B$. $\langle u\rangle$ is the average value of $u$ in the integral
in (\ref{SR:tree}) with $\langle u\rangle \approx 0.87$ for $q^2=0$
and $\langle u\rangle \approx 0.72$ for $q^2=15\,{\rm GeV}^2$.
Variations of the Borel parameter in the chosen window result in an
minimal uncertainty in the prediction for the form factor of about
10\%, independent of $q^2$. 
We are going to discuss the other remaining uncertainties in detail.

\begin{figure}
\centerline{\epsffile{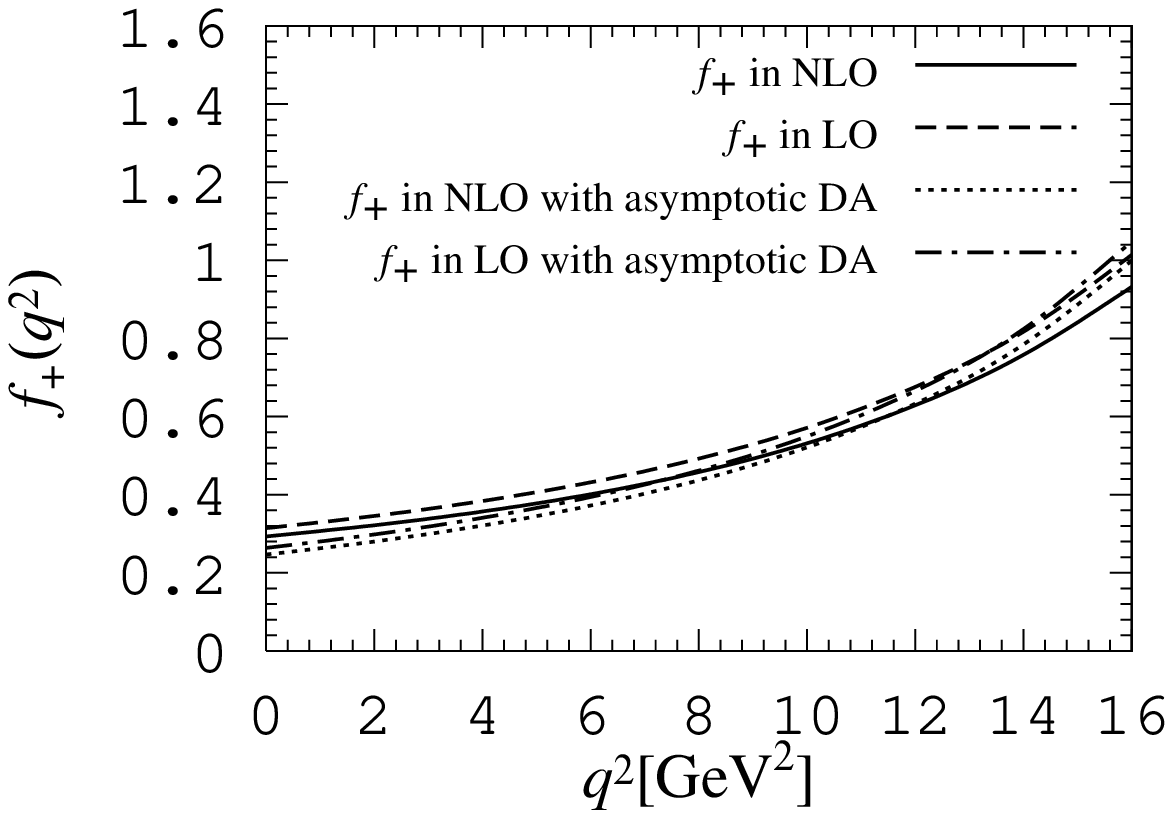}}
\caption[]{$f_+(q^2)$ as a function of $q^2$ for two
  different sets of the leading twist pion distribution amplitude. 
  The effect of including radiative corrections is a reduction of the
  form factor by about (4--7)\%.
}\label{fig:q2}
\centerline{\epsffile{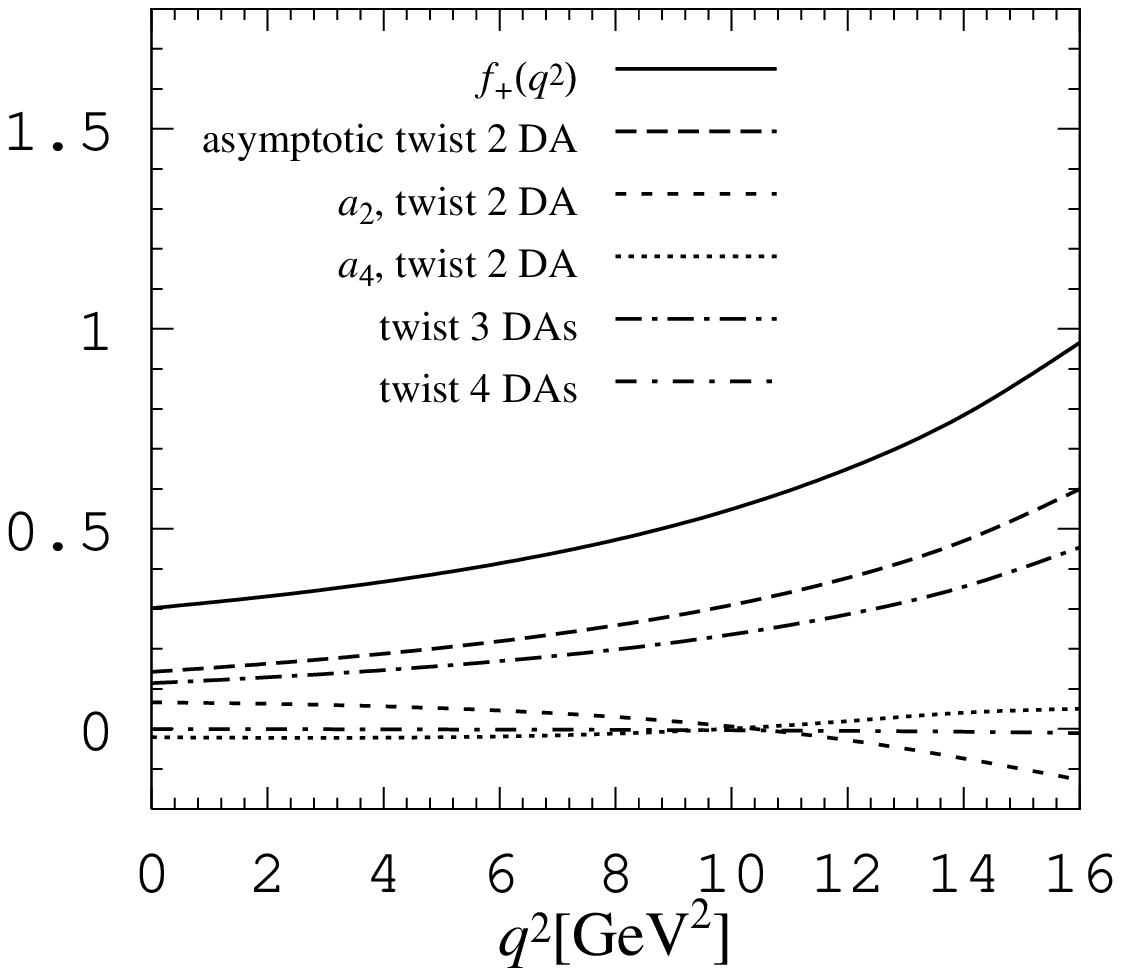}}
\caption[]{The several contributions to the light-cone sum rule for the form
  factor as a functions of $q^2$, using the leading twist distribution
  amplitude of {\protect\cite{BF1}}.}\label{fig:contr}
\end{figure}

{\it Radiative corrections.}\quad 
 Radiative corrections to the light-cone
 sum rule have been previously suspected 
 to be significant. It was expected,
 however, that large corrections to the correlation 
 function $\Pi_+$  partially cancel with  large corrections 
to the coupling $f_B$ in the ratio $f_+\sim \Pi_+/f_B$.
To quantify this effect, we write, schematically,
\begin{equation}
f_+ \sim \frac{\Pi_+^0 \left(1+\Pi_+^1 \,\frac{\alpha_s}{\pi}\right)}{
  f_B^0 \left( 1 + f_B^1\,\frac{\alpha_s}{\pi}\right)},
\label{SR:show}
\end{equation}
where $\Pi_+^0$ and $f_B^0$ are the tree-level contributions and
$\Pi_+^1$ and $f_B^1$ specify the corrections.
For central values of the input parameters, we find $f_B^1\simeq 3.0$ and
$\Pi_+^1=\{2.4, 2.3, 2.2, 2.1\}$\, for momentum transfers 
$q^2=\{0,5,10,15 \}\,$GeV$^2$, 
respectively. Thus the corrections indeed cancel 
each other to a large extent. The
``hard'' contribution to $\Pi_+^1$, defined by the terms in
(\ref{SR:rad})  involving an integral over $u$ in the interval
$0<u<u_0$, is $-0.5$ (at $q^2=0$), whereas the radiative correction to the
``soft'' contribution $u_0<u<1$ is $+2.9$. 
The resulting net effect of radiative corrections 
is shown in Fig.~\ref{fig:q2}, where we plot
$f_+$ as a function of $q^2$ for two different choices of the 
pion distribution amplitude (DA), see below.
The curves marked LO (NLO) are obtained by neglecting (including) 
radiative corrections in {\it both} the numerator and the denominator 
in (\ref{SR:show}).
The size of the correction is at most -7\%, thus providing 
 an {\em a posteriori} justification of the procedure of 
Refs.~\cite{BKR93,ABS,Bel95} to use a low value of $f_B$ in 
leading order light-cone sum rules.  

\begin{figure}
\centerline{\epsffile{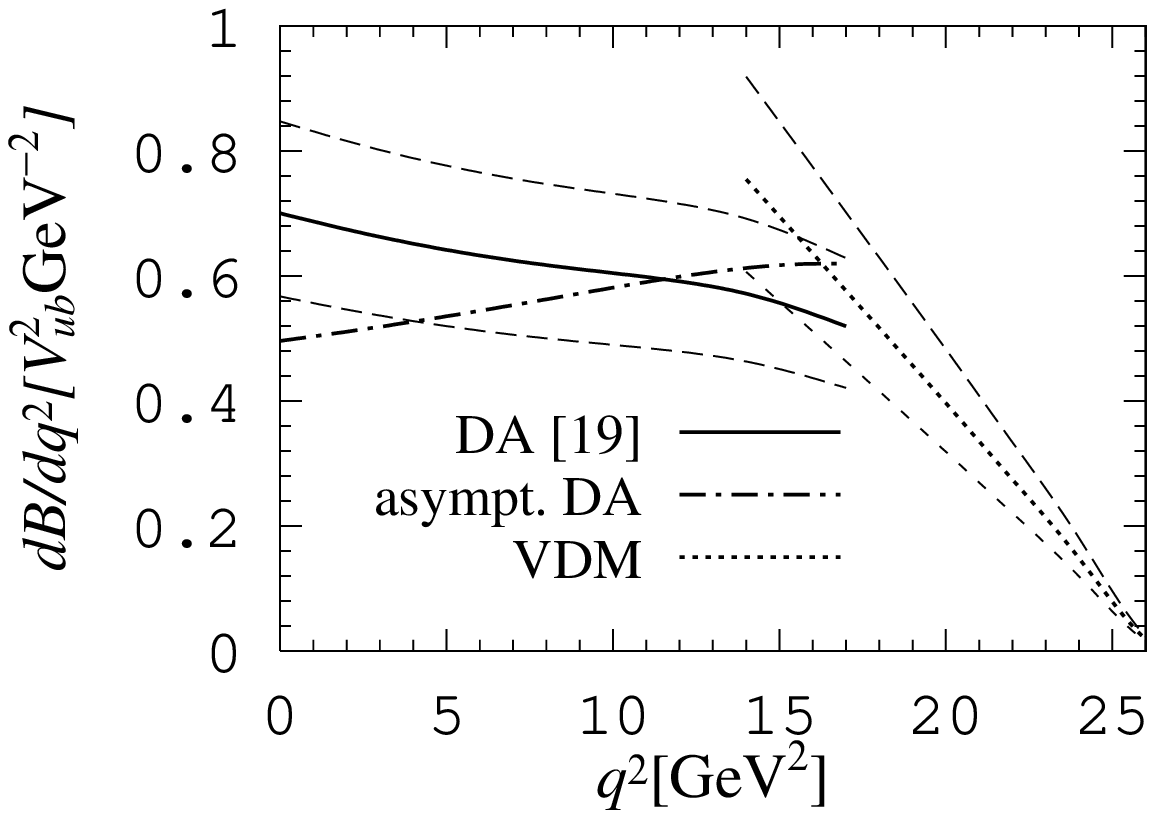}}
\caption[]{The spectrum $dB(B\to\pi e \nu)/dq^2$ as a function of
  $q^2$. Solid line: central values of the input parameters, using the
  distribution amplitude (DA)  \protect{\cite{BF1}}; 
  dashed-dotted line: the same using the
  asymptotic DA; dotted line: 
  the vector dominance approximation (VDM). The dashed
  lines indicate the range of the theoretical uncertainties. 
  See also text.
}\label{fig:spec}
\end{figure}

{\it Pion distribution amplitude.}\quad
The main
input in the sum rules is the pion distribution amplitude $\phi_\pi$
defined in
(\ref{def:phi_pi}),
which can conveniently be expanded in a series of Gegenbauer polynomials 
with multiplicatively renormalizable coefficients $a_n$ 
(to leading
logarithmic accuracy):\footnote{We neglect two-loop anomalous 
dimensions and also the
  mixing of the $a_n$, which occurs at two-loop accuracy. The corresponding
  expressions are available in the literature and can easily be
  incorporated in the analysis. Their effect is, however, negligible
  compared to uncertainties in the numerical values of the $a_n$.}
\begin{equation}
\phi_\pi(u,\mu) = 6u(1-u) \left[1+a_2(\mu) C_2^{3/2}(2u-1) + a_4(\mu)
  C_4^{3/2}(2u-1) +\ldots \right].
\end{equation}
The values of the nonperturbative coefficients $a_n$ can be restricted from 
experimental data, most notably from the
$\gamma\gamma^*\pi$ form factor. As argued in  \cite{rady}, the 
 recent  measurements  by CLEO \cite{CLEO} are consistent with the
distribution being close to its asymptotic form.
On the other hand, the old QCD sum rule results \cite{CZreport}
indicate sizable corrections which, in view of the criticism raised
in \cite{radmikh}, are probably overestimated.
In our analysis we use the QCD sum rule estimates
$a_2=0.35$ and $a_4=0.18$ at the  scale $\mu=2.4$~GeV
\cite{BF1,Bel95} as upper bounds 
for possible corrections to the asymptotic distributions.
The resulting $q^2$ dependence of the form factor is displayed
in Fig.~1 for the two choices --- asymptotic and QCD sum rule motivated ---
of the pion distribution amplitude. 
In Fig.~2 we show contributions of the asymptotic distribution 
and the corrections
separately. It is seen that
(a) the contribution of the second Gegenbauer polynomial $\sim a_2$ can
   reach 20\% and significantly affects  
   the shape of the form factor as a function of $q^2$ and that
(b) with current estimates of their magnitude, the contributions 
   of higher polynomials ($\sim a_4$) are negligible.  
We stress that the size of these corrections will eventually be
determined from experimental data.

{\it Higher twist corrections.}\quad
{}From Fig.~2 it is apparent that the twist three contribution to the sum rule
is large and of the same order as the leading twist contribution. 
An inspection shows that this large contribution is entirely due
to the asymptotic two-particle distribution
amplitudes of twist three, whose normalization is fixed by the quark 
condensate \cite{BF2}. 
A variation of $\langle \bar q q\rangle$ within the conservative 
limits $-(230-250$~MeV$)^3$ yields an uncertainty in $f_+(0)$ of at most
$\pm 0.012$, i.e.\ 4\%. 
Corrections to the asymptotic two-particle twist 3 distribution
amplitudes are related to the three-particle distribution of twist 3
 and proportional to the coupling $f_{3\pi}$ defined by the matrix element
\begin{equation}
\langle 0 | \bar d \sigma_{\mu\alpha}\gamma_5 G^\alpha_{\phantom{\alpha}\nu}
u | \pi^+(p)\rangle\ =\ 2 if_{3\pi} p_\mu p_\nu + \dots
\end{equation}   
The constant $f_{3\pi}$ was estimated from QCD sum rules to be
$f_{3\pi}= 0.0026\,$GeV$^2\pm$30\% (at the scale 2.4 GeV)
\cite{CZreport}.
The corresponding contribution to the form factor is within~2\%.
Finally, the twist 4 contributions  also turn out to be
unimportant numerically, see Fig.~2, so that 
we use the full set as specified in \cite{Bel95} 
without detailed error analysis. The only essential uncertainty in
the higher twist effects thus comes from the yet uncalculated
radiative corrections to the twist 3 contribution: a $+30$\% correction 
would increase the form factor by about 10\%.

Our final result for the spectrum 
$dB(B\to\pi e \nu)/dq^2$ as a function of
$q^2$ using the QCD sum rule motivated pion distribution amplitude 
and taking central values of the input parameters is shown 
in Fig.~3 (solid curve). 
We evaluate the light-cone sum rules for
$q^2<17\,$GeV$^2$, where both the twist-expansion and the contribution
of higher states are well under control. 
At higher values of $q^2$ the decay rate is strongly suppressed by the
phase space factor; for comparison, we show (dotted line) the spectrum calculated 
in the vector dominance (VDM) approximation
in this region, with the coupling $g_{BB^*\pi} =(29\pm 3) $ 
\cite{Bel95}. The uncentainties are illustrated by dashed lines.
The spectrum decreases monotonically with 
$q^2$, which is a consequence of the broad pion distribution
amplitude of \cite{BF1}. The asymptotic pion distribution
produces a different shape, see the dash-dotted curve, with a maximum 
around $q^2\sim 16-18$ GeV$^2$. We repeat that these two distributions 
present two extreme possibilities and the ambiguity will eventually
be removed. And vice versa, measuring the spectrum in $B\to\pi e\nu$
decays can distinguish between different shapes of the pion distribution
amplitude.    


To summarize, in this letter we have calculated the radiative correction
to the light-cone sum rule for the semileptonic $B\to\pi e\nu$ form factor.
We have studied its behaviour in the heavy quark limit,
which, as we believe, teaches us that
factorization of ``hard'' and ``soft''
subprocesses is generally valid in the heavy quark limit. 
{}From the numerical analysis 
of the sum rule, we obtain the form factor at $q^2=0$:
\begin{equation}
      f_+(0)  =  0.25\pm 0.03 \pm 0.01 \pm 0.01 
\end{equation}
assuming the asymptotic pion distribution amplitude and 
\begin{equation}
      f_+(0)  =  0.30\pm 0.03 \pm 0.01 \pm 0.01 
\end{equation}
using the QCD sum rule motivated distribution \protect{\cite{BF1}}.
The first error comes from the variation of the Borel parameter
within $5\,$GeV$^2<M^2<8\,$GeV$^2$, the second one from the error in
the quark condensate and the third one
from the combined uncertainties in $m_b$ and $f_{3\pi}$.
Combining all errors in quadrature and averaging over the two
different leading twist distributions, we get $f_+(0)=0.28\pm 0.05$ as
our final result.

We also give a simple parametrization of the $q^2$ dependence. 
For the asymptotic distribution we find:
\begin{equation}\label{para1}
f_+(q^2) = \frac{0.25\pm 0.03}{1- 1.72\,q^2/m_B^2 + 0.716\, q^4/m_B^4},
\end{equation}
and for the QCD sum rule distribution \cite{BF1}:
\begin{equation}\label{para2}
     f_+(q^2) = \frac{0.30\pm 0.03}{1- 1.32\,q^2/m_B^2 + 0.208\, q^4/m_B^4}.
\end{equation}
Both representations reproduce the exact light-cone sum rule results to within
1\% for $0<q^2<17\,$GeV$^2$. The influence of the other input
parameters on the $q^2$-dependence
is negligible, so we do not give errors in the denominators in
(\ref{para1}) and (\ref{para2}).

The total combined uncertainty of $\pm 0.05$
can be reduced to $\pm 0.03$ by calculating radiative
corrections to the asymptotic two-particle 
 distributions of twist~3 and from more detailed information on the 
pion distribution amplitude when it becomes available. 

\bigskip

\noindent {\bf Acknowledgements:} E.B. received support from CICYT research project AEN95-0815.
V.B.\ is grateful  to G. Korchemsky 
for the discussion of infrared factorization in
heavy quark decays and to the UAB and the CIRIT for financial support
during his stay in Barcelona. He also thanks the IFAE for kind hospitality.
P.B.\ acknowledges discussions with A. Khodjamirian, S. Weinzierl and 
O. Yakovlev. We thank M.~Lavelle for reading the manuscript.
 


\begin{thebibliography}{99}

\bibitem{ASY94} R.\ Akhoury, G.\ Sterman and Y.P.\ Yao,
 Phys.\ Rev.\ D {\bf 50} (1994) 358;\\
 R.\ Akhoury and I.Z.\ Rothstein, Phys.\ Lett.\ B {\bf 337} (1994) 176. 

\bibitem{hard}
A.\ Szczepaniak, E.M.\ Henley and S.J.\ Brodsky,
Phys.\ Lett.\ B {\bf 243} (1990) 287;\\
G.\ Burdman and J.P.\ Donoghue, Phys.\ Lett.\ B {\bf 270} (1991) 55.

\bibitem{EPStalk}V.M.\ Braun, 
talk given at Brussels EPS HEP (1995); published in  Brussels EPS HEP
(1995) 436 (hep--ph/9510404).

\bibitem{BKR93} V.M.\ Belyaev, A. Khodjamirian and  R. R\"{u}ckl, 
Z. Phys.\ C {\bf 60} (1993) 349;\\
A. Khodjamirian and  R. R\"{u}ckl, Talk presented at 28th
International Conference on High-energy Physics (ICHEP 96), Warsaw,
Poland, July 1996; published in  ICHEP 96, 902 (hep--ph/9610367).

\bibitem{Bel95} V.M.\ Belyaev et al., Phys.\ Rev.\ D {\bf 51} (1995) 6177.

\bibitem{KRWY} A.\ Khodjamirian et al., Preprint WUE-ITP-97-015 
(hep-ph/9706303). 

\bibitem{Braaten} E.\ Braaten, Phys.\ Rev.\ D {\bf 28} (1983) 524. 

\bibitem{exclusive} V.L.\ Chernyak and A.R.\ Zhitnitsky, JETP Lett.\
  {\bf {25}} (1977) 510; Yad.\ Fiz.\ {\bf 31} (1980) 1053;\\
  A.V.\ Efremov and A.V.\ Radyushkin, Phys.\ Lett.\ B {\bf 94} (1980)
  245;
  Teor.\ Mat.\ Fiz.\  {\bf {44}} (1980) 157;\\
  G.P.\ Lepage and S.J.\ Brodsky, Phys.\ Lett.\ B {\bf 87} (1979) 359;
  Phys.\ Rev.\ D {\bf 22} (1980) 2157.

\bibitem{CZ90} V.L.\ Chernyak and I.R.\ Zhitnitsky, Nucl.\ Phys.\ {\bf
    B345} (1990) 137.

\bibitem{ABS} A. Ali, V.M.\ Braun and H. Simma, Z.\ Phys.\ C {\bf 63}
  (1994) 437.


\bibitem{KS94} 
G.P.\ Korchemsky and G. Sterman, Phys.\ Lett.\ B {\bf 340} (1994) 96.

\bibitem{BBBD} E. Bagan et al., Phys.\ Lett.\ B {\bf {278}} (1992) 457.

\bibitem{AE} T.M.\ Aliev and V.L.\ Eletskii, Yad.\ Fiz.\ {\bf 38}
  (1983) 1537.
 
\bibitem{Flynn96} J. Flynn, Talk given at Lattice 96,
14th International Symposium on Lattice Field Theory, St.\ Louis, USA,
June 1996; 
published in Nucl.\ Phys.\ Proc.\ Suppl.\ {\bf 53} (1997) 168
(hep--lat/9610010). 

\bibitem{rady} I.V.\ Musatov and A.V.\ Radyushkin, Phys.\ Rev.\ D {\bf
  56} (1997) 2713.

\bibitem{CLEO} 
J. Gronberg et al.\ (CLEO coll.), Preprint CLNS-97-1477 (hep-ex/9707031). 

\bibitem{CZreport} V.L.\ Chernyak and I.R.\ Zhitnitsky, Phys.\ Rep.\
  {\bf 112} (1984) 173.

\bibitem{radmikh} S.V.\ Mikhailov and A.V. Radyushkin, Phys.\ Rev.\ D
{\bf 45} (1992) 1754.

\bibitem{BF1} V.M.\ Braun and I.E.\ Filyanov, Z.\ Phys.\ C {\bf {44}}
  (1989) 157.

\bibitem{BF2} V.M.\ Braun and I.E.\ Filyanov, Z.\ Phys.\ C {\bf {48}}
  (1990) 239.

\end{thebibliography}
\end{document}